\documentstyle[11pt,epsfig]{article}

\def\laq{~\raise 0.4ex\hbox{$<$}\kern -0.8em\lower 0.62
ex\hbox{$\sim$}~}
\def\gaq{~\raise 0.4ex\hbox{$>$}\kern -0.7em\lower 0.62
ex\hbox{$\sim$}~}

\begin{document}

\begin{titlepage}
\begin{flushright}
CERN-PH-TH/2010-120
\end{flushright}
\vspace*{1.5 cm}

\begin{center}
\huge{Multiplicity distributions\\
in gravitational and strong interactions}
\vskip 0.5cm
\large{Massimo Giovannini\footnote{e-mail address: massimo.giovannini@cern.ch}}
\vskip 0.5cm
{\it   Department of Physics, Theory Division, CERN, 1211 Geneva 23, Switzerland}
\vskip 0.5cm
{\it  INFN, Section of Milan-Bicocca, 20126 Milan, Italy}
\vskip 1cm

\begin{abstract}
The multiplicity distributions produced by the variation of time-dependent  
gravitational fields in a conformally flat background geometry
 belong to the same class of infinitely divisible distributions found, for fixed centre of mass energies and symmetric (pseudo)rapidity intervals, in charged multiplicities produced in $pp$, $p\overline{p}$ and in heavy ion collisions.  
Apparently unrelated multiplicity distributions are classified in terms of the (positive) discrete representations of the $SU(1,1)$ group. The gravitational 
analogy suggest a global high-energy asymptote for the distributions measured in $pp$ and $p\overline{p}$ collisions. 
Second-order cross correlations between positively and negatively charged distributions represent a relevant diagnostic for a closer 
scrutiny of the multiparticle final state.
\end{abstract}
\end{center}
\end{titlepage}

\newpage
A. Sakharov \cite{sakharov} was presumably the first to raise the question of the quantum mechanical 
origin of density perturbations in the early Universe suggesting that 
the complicated patterns observed in the galaxy distributions could have a plausible origin in the zero-point fluctuations of 
matter and radiation fields in curved backgrounds. More recently the latter 
perspective gained  a firmer support from the analyses of the Cosmic Microwave 
Background (CMB) anisotropies and polarization. It can therefore be speculated that 
the initial conditions of the CMB  anisotropies were actually set in a regime 
where the expansion rate was $H \simeq {\mathcal O}(10^{-6} M_{\mathrm{P}}) \simeq {\mathcal O}(10^{37}
\mathrm{Hz})$. Today the background geometry is, to a good approximation, conformally flat at least according to the WMAP data in their various releases \cite{WMAP} (see also, for instance, \cite{weinberg}) and in the light 
of the so-called $\Lambda$CDM lore where $\Lambda$ stands for the cosmological constant and CDM for the cold dark matter component.

Gravitational and strong interactions are intrinsically different in the physical regimes where direct laboratory  tests are possible.  
Even acknowledging, for large space-time curvatures, the corresponding largeness of the expansion rate, multiple production in strong gravitational fields is naively expected to be drastically different from the laws governing the dynamics of the multiparticle final state in hadronic processes such as, for instance, 
proton-proton (i.e. $pp$), proton-antiproton (i.e. $p\overline{p}$) or even heavy ions collisions. The purpose of the present analysis is to challenge 
the latter statement by analyzing the multiple production of electrically charged species (i.e. positive and negative) in conformally flat geometries and in a  correct dynamical 
framework. Consider therefore  one of the simplest models for the production of charged species, i.e. a minimally coupled complex scalar field in a conformally flat geometry; the relevant four-dimensional action can be written 
\begin{equation}
S= \int d^{4} x \, \sqrt{-g} \, \biggl[ g^{\mu \nu} \partial_{\mu} \phi^{*} \partial_{\nu} \phi 
- m^2 \phi^{*} \phi\biggr],
\label{act1}
\end{equation}
where $g_{\mu\nu} = a^{2}(\tau) \eta_{\mu\nu}$ is the (conformally flat) background metric of Friedmann-Robertson-Walker type expressed in terms of the conformal time coordinate $\tau$ and in terms of the scale factor $a(\tau)$; 
$\eta_{\mu\nu}$ denotes the Minkowski metric with signature $(+,\,-,\,-,\, -)$.
By introducing the canonical normal modes $\Phi(\vec{x},\tau) = a(\tau) \phi(\vec{x}, \tau)$ and $\Phi^{*}(\vec{x},\tau) = a(\tau) 
\phi^{*}(\vec{x}, \tau)$ the canonical Hamiltonian can be written as
\begin{equation}
H(\tau) = \int d^{3} x \biggl[ \Pi^{*} \Pi + {\mathcal H}(\Phi\, \Pi + 
\Phi^{*} \Pi^{*}) + \partial_{i} \Phi^{*} \partial^{i} \Phi + m^2 a^2 \Phi^{*} \Phi\biggr],
\label{act4}
\end{equation}
where ${\mathcal H} = \partial_{\tau} (\ln{a})$ denotes 
the derivative of the natural logarithm of the scale factor with respect to the conformal time coordinate and $\Pi = \partial_{\tau} \Phi^{*} - {\mathcal H} \Phi^{*}$ 
is the canonical momentum.  By promoting the classical fields to quantum mechanical 
operators obeying (equal time) commutation relations the Hamiltonian (\ref{act4}) 
can be written as the sum of a free part and of an interacting part $\hat{H}(\tau) = \hat{H}_{0}(\tau) + \hat{H}_{\mathrm{I}}(\tau)$ with $\hat{H}_{0}(\tau)$ and 
$\hat{H}_{\mathrm{I}}(\tau)$ given, respectively, by 
\begin{equation}
\hat{H}_{0}(\tau) =2  \int d^{3} p \,\,\omega(\tau)\,\, K_{0}(\vec{p}) , \qquad \hat{H}_{\mathrm{I}}(\tau) =  2\int\,\,d^{3} p\,\, [ \lambda^{*}(\tau) K_{-}(\vec{p}) + \lambda(\tau) K_{+}(\vec{p})],
\label{act9}
\end{equation}
where $\omega(\tau) = \sqrt{p^2 + m^2 a^2}$, $\vec{p}$ is the 
comoving three-momentum and $\lambda(\tau) = i {\mathcal H}/2$;
 the operators $K_{\pm}(\vec{p})$ and $K_{0}(\vec{p})$ are defined as:
\begin{equation}
 K_{+}(\vec{p}) = \hat{a}_{\vec{p}}^{\dagger} \,\hat{b}_{-\vec{p}}^{\dagger},\qquad K_{-}(\vec{p}) = \hat{a}_{\vec{p}}\, \hat{b}_{-\vec{p}},
\qquad K_{0}(\vec{p}) = \frac{1}{2}\biggl[ \hat{a}_{\vec{p}}^{\dagger}\, \hat{a}_{\vec{p}} + \hat{b}_{-\vec{p}} \,\hat{b}_{-\vec{p}}^{\dagger}\biggr].
\label{act10}
\end{equation}
Since  $[\hat{a}_{\vec{p}}, \hat{a}^{\dagger}_{\vec{k}}] = 
\delta^{(3)}(\vec{p} - \vec{k})$, $[\hat{b}_{\vec{p}}, \hat{b}^{\dagger}_{\vec{k}}] = 
\delta^{(3)}(\vec{p} - \vec{k})$ and $[\hat{a}_{\vec{p}}, \hat{b}_{\vec{k}}]=0$, the operators (\ref{act10}) satisfy the commutation relations
\begin{equation}
 [K_{-}(\vec{p}) , K_{+}(\vec{q})] = 2 \,K_{0}(\vec{p}) \,
\delta^{(3)}(\vec{p} - \vec{q}),\qquad [K_{0}(\vec{p}), K_{\pm}(\vec{q})] = \pm \,K_{\pm}(\vec{p})\, \delta^{(3)}(\vec{p}-\vec{q}),
\label{act11}
\end{equation}
showing that $K_{\pm}(\vec{p})$ and $K_{0}(\vec{p})$ are nothing but the generators the $SU(1,1)$ group obeying the commutation relations 
of the corresponding Lie algebra.
Owing to the group structure (\ref{act11}) and to the specific form of the 
Hamiltonian of Eq. (\ref{act9}), the multiparticle  state for $\tau\to +\infty$ can be obtained by applying 
to the initial state  $|\Psi_{i}(\vec{p})\rangle$ the product of two unitary operators ${\mathcal R}(\varphi_{p})$ and $\Sigma(z_{p})$:
\begin{equation}
|\Psi_{f}(\vec{p})\rangle = {\mathcal R}(\varphi_{p}) \, \Sigma(z_{p}) |\Psi_{i}(\vec{p})\rangle, \qquad |\Psi_{f}\rangle = \prod_{\vec{p}} |\Psi_{f}(\vec{p})\rangle,
\label{act12}
\end{equation}
where $|\Psi_{f}(\vec{p})\rangle$ denotes the final state and where the unitary operators are given by:
\begin{equation}
{\mathcal R}(\varphi_{p}) = \exp{[ - 2 i \varphi_{p}\, K_{0}(\vec{p})]}, \qquad \Sigma(z_{p}) = \exp{[z_{p}^{*}\, K_{-}(\vec{p}) - 
z_{p}\, K_{+}(\vec{p})]},
\label{act13}
\end{equation}
with $z_{p} = r_{p} e^{i \vartheta_{p}}$ and $\alpha_{p} = (2\varphi_{p} -\vartheta_{p})$. Denoting with a an overdot a derivation 
with respect to $\tau$, 
the time evolution of the 
variables $r_{p}(\tau)$, $\varphi_{p}(\tau)$  and $\alpha_{p}(\tau)$ is given by 
\footnote{It should be noted that the factorization of the time-evolution 
operator in terms of ${\mathcal R}(\varphi_{p})$ and of $\Sigma(z_{p})$ is non-trivial since $\hat{H}_{0}$ and $\hat{H}_{\mathrm{I}}$ 
do not commute. An analog problem arises  in two-photon optics \cite{sch1,gl}.}  $\dot{r}_{p} = - {\mathcal H} \cos{\alpha_{p}}$, 
$\dot{\varphi}_{p} = \omega + {\mathcal H} \tanh{r_{p}} \sin{\alpha_{p}}$, and $\dot{\alpha}_{p}
= 2 \omega + 2 {\mathcal H} \sin{\alpha_{p}}/\tanh{2 r_{p}}$.
The explicit solution of the latter equations depends upon the evolution of ${\mathcal H}$ which is determined, ultimately, 
by the evolution of the space-time curvature. The purpose here will not be to compute the average multiplicity in a given 
model but rather to analyze the correlation properties of  the multiplicity distributions.

The initial quantum state for each mode of the field (i.e. $|\Psi_{i}(\vec{p}) \rangle$)  
can be classified in terms of the Fock basis $|n_{+}\,\, n_{-} \rangle$ which is
 an appropriate basis for the irreducible representations of $SU(1,1)$ once 
the group generators are represented as in Eq. (\ref{act10}) (see also \cite{schwinger} for the $SU(2)$ case). An equivalent basis for the irreducible representations of $SU(1,1)$ is provided by the vectors  $|q\,n_{\mathrm{ch}}\rangle$ where 
$q = n_{+} - n_{-}$ is the total charge and $n_{\mathrm{ch}} = n_{+} + n_{-}$
is the total number of charged species. 
The vectors  $|q\,n_{\mathrm{ch}}\rangle$  are the standard basis of the irreducible representations $T^{+k}$ of $SU(1,1)$ where $k$ is the principal quantum number and $m$ is the magnetic quantum number, i.e. the eigenvalue of $K_{0}$. 
The negative series $T^{-k}$ is symmetric under the exchange $n_{+} \to n_{-}$ while the principal (continuous) series will not play a specific role in the present considerations. In terms of $k$ and $m$ we have that the total charge and the total number of particles are given, respectively, by 
$q= 2 k -1$ and by $n_{\mathrm{ch}} = 2m -1$. Since $2k= {\mathcal N}$ is a positive integer, the charge and the total number of charged species will be, respectively,  $q = {\mathcal N} -1 = 0,\,1,\,2,\,...$ and $n_{\mathrm{ch}}= q, \, q+ 1,\, q+ 2...$ and so on.  
According to the Backer-Campbell-Hausdorff (BCH) decomposition, the operator of Eq. (\ref{act13}) can be 
factorized as the product of the exponentials of the group generators  \cite{sch1,pere}, i.e. 
$\Sigma(z_{p}) = {\mathcal A}_{+}(z_{p}){\mathcal A}_{0}(z_{p}){\mathcal A}_{-}(z_{p})$ where ${\mathcal A}_{0}(z_{p}) = \exp{[ - 2 \ln{(\cosh{r_{p}})} K_{0}(\vec{p})]}$, ${\mathcal A}_{-}(z_{p})  = \exp{[ e^{- i \vartheta_{p}} \tanh{r_{p}} \, K_{-}(\vec{p})]}$,
and ${\mathcal A}_{+}(z_{p}) = \exp{[ - e^{ i \vartheta_{p}} \tanh{r_{p}} \, K_{+}(\vec{p})]}$. Using the BCH decomposition, as well as the explicit form of $|\Psi_{f}(\vec{p})\rangle$ given in Eq. (\ref{act12}), the density operator can be deduced. Defining the following triplet of functions
\begin{eqnarray}
&& C(m, \ell, j) = \frac{\Gamma(m +1)\, \Gamma(m - \ell + j +1)}{\Gamma(\ell +1) \, \Gamma(j+1) \Gamma^2(m - \ell +1)},
\label{dm1}\\
&& {\mathcal M}(n_{\pm}, \ell, j, \ell', j')= \sqrt{C(n_{+}, \ell, j)C(n_{-}, \ell, j)C(n_{+}, \ell', j')C(n_{-}, \ell', j')},
\label{dm2}\\
&& {\mathcal F}( r_{p}, \alpha_{p}; n_{\pm}; \ell, \ell'; j,j' ) = e^{- i \alpha_{p}[(j - \ell) - (j' - \ell')]} \frac{(-\tanh{r_{p}})^{j + j' + \ell + \ell'}}{(\cosh^2r_{p})^{n_{+} + n_{-} - \ell -\ell' +1 }},
\label{dm3}
\end{eqnarray}
the density operator reads 
\begin{eqnarray}
&& \hat{\rho}_{f}(p) = \sum_{j,\,j'=0}^{\infty} \,\,\sum_{\ell,\,\ell' =0}^{\ell_{\mathrm{max}}}   {\mathcal G}(n_{\pm};\,\ell,\ell';\, j,j')
| n_{+} -\ell+j,\, n_{-} -\ell + j\rangle \langle j' - \ell' + n_{-},\, j -\ell + n_{+} |,
\label{dm4}\\
&&{\mathcal G}(n_{\pm};\,\ell,\ell';\, j,j') = {\mathcal F}( r_{p},\alpha_{p}; n_{\pm}; \ell, \ell'; j,j' ) 
{\mathcal M}(n_{\pm},\ell, j, \ell', j'),
\label{dm5}
\end{eqnarray}
where $\ell_{\mathrm{max}} = {\mathrm Min}(n_{+},\,n_{-})$. In terms of the density matrix of Eq. (\ref{dm4}) the expectation value of a generic operator $\hat{O}$ can be 
computed as $\langle \hat{O} \rangle = \mathrm{Tr}[ \hat{\rho}\, \hat{O}]$. Consider first, in Eq. (\ref{dm4}), the case $q = n_{\mathrm{ch}} =0$.  Since, in this case, 
 $\ell_{\mathrm{max}} =0$, Eq. (\ref{dm4}) implies:
\begin{equation}
\hat{\rho}_{f}(p) = \frac{1}{\overline{n} +1} \sum_{m,\,n=0}^{\infty} (-1)^{m + n} e^{- i \alpha_{p}(m-n)} \biggl(\frac{\overline{n}}{\overline{n} +1}\biggr)^{(m + n)/2} | m\, m\rangle \langle n\,n|,
\label{av6}
\end{equation}
where $\langle \hat{N}_{+} + \hat{N}_{-} \rangle = 2 \sinh^2{r_{p}} = 2 \overline{n}$ and $\overline{n}$ denotes the average multiplicity of pairs. The density matrix of Eq. (\ref{av6}) is idempotent (i.e.  $\hat{\rho}_{f}^2 = \hat{\rho}_{f}$) but, nonetheless the diagonal elements of $\hat{\rho}_{f}(\vec{p})$  follow a Bose-Einstein (BE) multiplicity distribution: 
\begin{equation}
P_{n}^{(\mathrm{BE})}(\overline{n}) = \langle n\,n| \hat{\rho}_{f}(p)|n\, n \rangle=  \frac{\overline{n}^{n}}{(\overline{n} +1)^{n+1}}, \qquad \sum_{n=0}^{\infty} P_{n}^{(\mathrm{BE})}(\overline{n})=1.
\label{av7}
\end{equation}
The integration of Eq. (\ref{av6}) over $\alpha_{p}$ between $0$ and $2 \pi$  defines the reduced density operator in the random phase approximation: 
the mixed state obtained with this procedure will have a thermal density matrix. This simple observation shows that a BE 
 multiplicity distribution is not sufficient to infer local thermal equilibrium: as an example recall that 
chaotic light distributed as BE can be generated by sources in which atoms are kept at an excitation level higher than that in thermal equilibrium \cite{loudon2}. The multiplicity distributions of Eqs. (\ref{av6}) and (\ref{av7}) are also typical of the two-mode squeezed vacuum states
\cite{sch1,mandel,knight}. The two-mode 
squeezed vacuum state of photons can be used to describe multigraviton \cite{cossq1}, 
multiphonon \cite{cossq2}, and multiphoton \cite{cossq3} 
states in the quantum treatment of cosmological inhomogeneities (see also \cite{mgbook} for an introductory discussion).  
Consider next the case $q = n_{\mathrm{ch}} = {\mathcal N}-1$, i.e. the situation in which the initial state 
has a definite charge. Equation (\ref{dm4}) implies that the diagonal elements of the density operator can be written as
\begin{equation}
 P_{n}(\overline{n}, {\mathcal N}) = \frac{\Gamma(n + {\mathcal N})}{\Gamma({\mathcal N}) \Gamma(n+1)} \frac{\biggl(\frac{\overline{n}}{{\mathcal N}} - \epsilon \biggr)^{n}}{\biggl( \frac{\overline{n}}{{\mathcal N}} +  1 -\epsilon\biggr)^{n + {\mathcal N}}}, \qquad 
 \epsilon = \frac{{\mathcal N}-1}{2 {\mathcal N}},
\label{av12}
\end{equation}
which reduces to the second expression of Eq. (\ref{av7}) when ${\mathcal N} \to 1$.  The multiplicity distribution of Eq. (\ref{av12}) falls into the 
class of negative binomial multiplicity distributions. Indeed,  by expanding Eq. (\ref{av12}) in powers of $\epsilon = ({\mathcal N} -1)/(2{\mathcal N}) < 1$:
\begin{eqnarray}
&&P_{n}(\overline{n}, {\mathcal N}) = P^{(\mathrm{NB})}_{n}(\overline{n}, {\mathcal N})\biggl[ 1 + (\overline{n} - n) \biggl( \frac{{\mathcal N}}{\overline{n}+ {\mathcal N}}\biggr)\biggl(\frac{{\mathcal N}}{\overline{n}}\biggr) \epsilon + {\mathcal O}(\epsilon^2) \biggl],
\label{av12a}\\
&& P_{n}(\overline{n}, k_{\mathrm{NB}}) = \frac{\Gamma(n + k_{\mathrm{NB}})}{\Gamma(k_{\mathrm{NB}}) \Gamma(n+1)} 
\biggl(\frac{\overline{n}}{\overline{n} + k_{\mathrm{NB}}}\biggr)^{n} 
 \biggl(\frac{k_{\mathrm{NB}}}{\overline{n} + k_{\mathrm{NB}}}\biggr)^{k_{\mathrm{NB}}},
 \label{av13}
 \end{eqnarray}
 where $P_{n}(\overline{n}, k_{\mathrm{NB}})$ denotes the negative binomial (NB) probability distribution with parameters $\overline{n}$ and $k_{\mathrm{NB}}$. 
Equation (\ref{av12}) is in fact a distorted negative binomial distribution (DNB): according 
to Eq. (\ref{av12a})  $P_{n}(\overline{n}, {\mathcal N})$ is larger than $P_{n}(\overline{n},{\mathcal N})$ 
for $n < \overline{n}$ and it is smaller than $P_{n}(\overline{n}, {\mathcal N})$ for $n > \overline{n}$.  
\begin{figure}
\begin{center}
\begin{tabular}{|c|c|}
      \hline
      \hbox{\epsfxsize = 7.2 cm  \epsffile{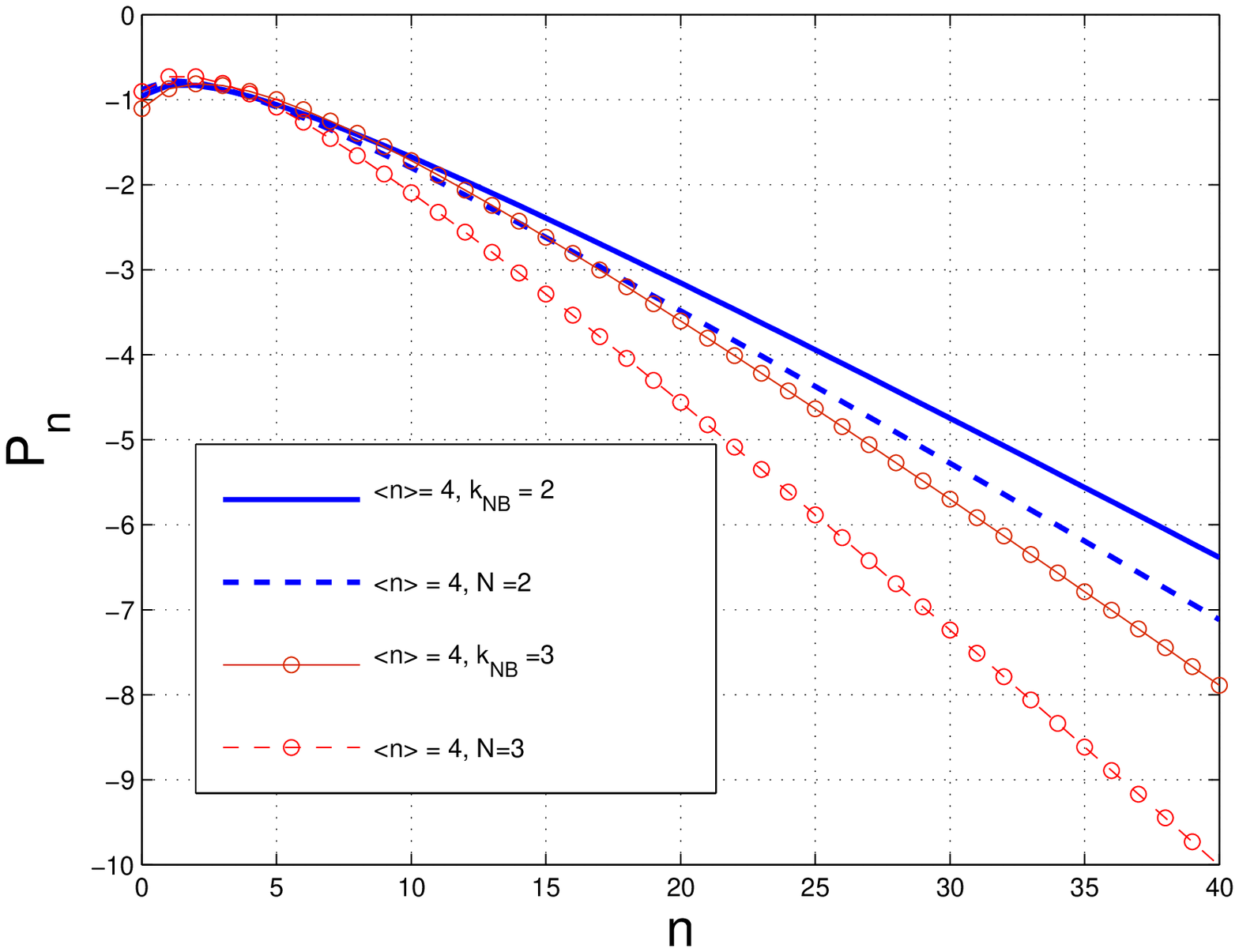}} &
      \hbox{\epsfxsize = 7.2 cm  \epsffile{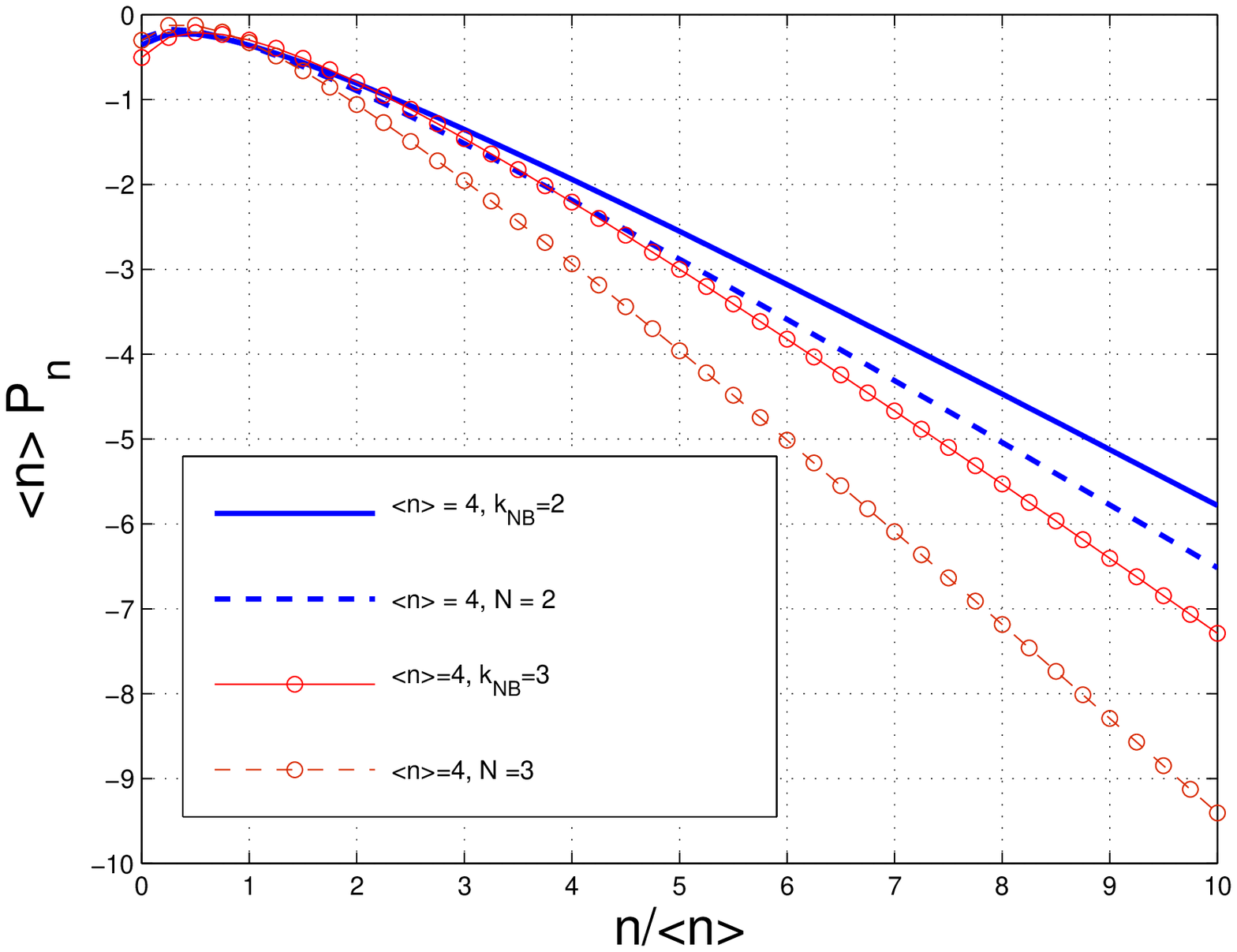}}\\
      \hline
\end{tabular}
\end{center}
\caption{The distributions of Eqs. (\ref{av12}) and (\ref{av13}) are compared for the same range of parameters. In both  plots the 
scale is linear on the horizontal axis, while, on the vertical axis the common logarithm is reported. In the legends of both plots $\langle n \rangle$ coincides with the $\overline{n}$ appearing in Eqs. (\ref{av12}) and (\ref{av13}). 
The naughts on the two (lower) thin lines are not experimental points 
but just a graphic means to distinguish different curves reminding visually that we are dealing here with discrete 
distributions. } 
\label{FIGURE1}
\end{figure}
In Fig. \ref{FIGURE1} the DNB distribution of Eq. (\ref{av12}) is compared, for the same 
range of parameters, with the standard form of the NB distribution 
(see Eq. (\ref{av13})). The 
full lines denotes the standard negative binomial result while the dashed lines 
refer to the DNB of Eq. (\ref{av12}) which undershoots (in comparison with the standard NB case) 
when the multiplicity exceeds $\overline{n}$ and overshoots in the opposite case, i.e. when $n < \overline{n}$. In the plot at the right of Fig. \ref{FIGURE1} the notation is the same but the distributions are illustrated in Koba-Nielsen-Olesen (KNO for short) variables \cite{KNO}.
While the KNO scaling is violated since $k_{\mathrm{NB}}$ changes with the centre of mass energy of the collision (see below), there is still the useful habit to present the results for the multiplicity distributions in terms of KNO variables.  
To complete the discussion of the limits of Eq. (\ref{dm4}) it is interesting to mention, the case when
 $n_{\mathrm{ch}} \neq 0$ but $q=0$; the density matrix of Eq. (\ref{dm4}) leads then to a more complicated multiplicity distribution whose explicit form can be written as 
\begin{equation}
P_{m}(n,x) = \frac{{x}^{m -n}}{(x +1)^{m + n +1}} \Gamma^2(m + 1) \Gamma^2(n +1)\biggl[ \sum_{\ell =0}^{n} \frac{(-x)^{\ell}}{\Gamma(m - n + \ell +1) \Gamma(\ell +1) \Gamma^2(n -\ell +1)} \biggr]^2,
\label{av15}
\end{equation}
where $x= \sinh^2{r_{p}} = (2 \overline{n} - n)/[2 ( n +1)]$. In the case $n = 0$ we have that $\ell_{max}=0$: the finite 
sum appearing in Eq. (\ref{av15}) gives, after squaring, $1/\Gamma^2(m+1)$ and $P_{m}(x) \to P_{m}^{(\mathrm{BE})}$ 
with $x = \overline{n}$. The multiplicity distributions described by Eq. (\ref{av15}) have several interesting features 
like oscillations in $m$ for fixed values of $\overline{n}$ and $n$. These properties will not be specifically discussed here.  Neglecting some phases, the connection between the multiplicity distributions and the $SU(1,1)$ group structure can be neatly expressed by computing, in explicit terms, the Wigner matrix element of the positive discrete series:
\begin{eqnarray}
T^{+k}_{m\,\,m'} &=& \langle k\,m' | \Sigma(z_{p}) |k m \rangle =   \frac{\sqrt{ \Gamma(k + m') \,\, \Gamma(m' - k+1)}}{\sqrt{ \Gamma(k + m)\,\, \Gamma(m - k+1)}} 
\frac{( - e^{ i \vartheta_{p}} \tanh{r_{p}})^{m' - m} }{ (m' - m)!\,\, (\cosh{r_{p}})^{2 m}} 
\nonumber\\
&\times& F[ 1 - m - k,\, k - m;\, m' - m +1; -\sinh^2{r_{p}}], \qquad m'> m,
\label{av20}
\end{eqnarray}
where $F[\alpha,\beta;\gamma; x]$ is the hypergeometric function. 
From Eq. (\ref{av20}) we have, immediately, that 
\begin{eqnarray}
&& \biggl| T^{1/2}_{1/2\,\, n+ 1/2}\biggr|^2 = \frac{1}{\overline{n} + 1} \biggl(\frac{\overline{n}}{\overline{n} +1}\biggr)^{n} \equiv P_{n}^{(\mathrm{BE})}(\overline{n}) , \qquad 
\overline{n} = \sinh^2{r_{p}},
\label{av21}\\
&&  \biggl| T^{\ell/2}_{\ell/2\,\, n +\ell/2}\biggr|^2 = \frac{\Gamma(\ell + n)}{\Gamma(n+1) \Gamma(\ell)} \biggl(\frac{\overline{n}}{\overline{n} +\ell}\biggr)^{n} \biggl(\frac{\ell}{\ell + \overline{n}}\biggr)^{\ell}\equiv P_{n}^{(\mathrm{NB})}(\overline{n},\ell), \qquad \sinh^2{r_{p}} = \frac{\overline{n}}{\ell}.
\label{av22}
\end{eqnarray} 
The multiplicity distribution $P_{n}(\overline{n}, k_{\mathrm{NB}})$ has various notable limits. Indeed, by computing the corresponding probability generating function (pgf), 
\begin{equation}
 {\mathcal P}(z, \overline{n}, k_{\mathrm{NB}}) = \sum_{k=0}^{\infty} z^{k} P_{k}(\overline{n},k_{\mathrm{NB}}) = 
 \frac{{k_{\mathrm{NB}}}^{k_{\mathrm{NB}}}}{[\overline{n}(1 - z) + k_{\mathrm{NB}}]^{k_{\mathrm{NB}}}},
 \label{av13b}
 \end{equation}
it is immediate to see that, in the limit $k_{\mathrm{NB}} \to 1$ we recover 
the pgf of the Bose-Einstein distribution while, in the limit $k_{\mathrm{NB}} \to \infty$ (and $\overline{n}$ fixed) 
${\mathcal P}(z,\overline{n}, k_{\mathrm{NB}}) \to \exp{[(z -1)\overline{n}]}$, i.e. 
the pgf of the Poisson distribution. Finally, both the Gamma as well as the logarithmic 
distributions can be obtained as specific limits of the NB distribution.
The distributions arising as matrix elements of $SU(1,1)$ from the recipe of Eqs. (\ref{av21}) and (\ref{av22}) 
are all infinitely divisible\footnote{A probability distribution is said infinitely divisible if, for any 
given non negative integer $k$, it is possible to find $k$ independent identically distributed random variables whose probability distributions sum up to the original distribution. The generating function of the sum of independent identically distributed random variables is given by the product of the generating functions of each distribution of the sum. On the basis of the latter theorem, a distribution with
probability generating function ${\mathcal P}(z)$ is infinitely divisible provided, for any integer $k$, there exist $k$ independent identically distributed random variables with generating function ${\mathcal Q}_{k}(z)$ such that ${\mathcal P}(z) = [{\mathcal Q}_{k}(z)]^{k}$.}. While a more thorough 
investigation of the latter statement is beyond the scopes of the present discussion, it is suggestive to note that neither 
the binomial distribution nor the uniform distribution (which are not infinitely divisible) 
arise in the proposed correspondence between $SU(1,1)$ matrix elements and the discrete 
multiplicity distributions.  

Multiplicity distributions in general (and the gravitational multiplicity distributions in particular)  can be classified by using the degree of second order coherence. Borrowing the terminology of quantum optics, multiplicity distributions can be classified according to the their normalized two-point function \cite{loudon2,mandel}
\begin{eqnarray}
&& g_{+}^{(2)} -1 = \frac{\langle  \hat{a}^{\dagger}_{\vec{p}} \,  \hat{a}^{\dagger}_{\vec{p}}\, \hat{a}_{\vec{p}}\hat{a}_{\vec{p}},\rangle  - \langle \hat{N}_{+}\rangle^2}{\langle \hat{N}_{+}\rangle^2},
\qquad g_{-}^{(2)} -1 = \frac{ \langle  \hat{b}^{\dagger}_{-\vec{p}} \,  \hat{b}^{\dagger}_{-\vec{p}}\, \hat{b}_{-\vec{p}}\,\hat{b}_{-\vec{p}}\rangle - \langle \hat{N}_{-}\rangle^2}{\langle \hat{N}_{-}\rangle^2}, 
\label{av4}\\
&&  g_{\pm}^{(2)} -1 = \frac{ \langle \hat{a}^{\dagger}_{\vec{p}} \, \hat{b}^{\dagger}_{-\vec{p}} \, \hat{a}_{\vec{p}}\, \hat{b}_{-\vec{p}}\rangle - \langle \hat{N}_{+}\rangle\langle N_{-}\rangle }{\langle \hat{N}_{-}\rangle\langle N_{+}\rangle}.
\label{av5}
\end{eqnarray}
If $q = n_{\mathrm{ch}} = 0$, Eqs. (\ref{av4}) and (\ref{av5}) imply  
$g_{+}^{(2)} = g_{-}^{(2)} =2$ while $g^{(2)}_{\pm}= 2 + 1/\overline{n}$,
where, as in Eq. (\ref{av7}), $\overline{n} = \sinh^2{r_{p}}$ denotes the averaged multiplicity of pairs.
If $q = 0$ but $n_{\mathrm{ch}} \neq 0$, $g^{(2)}_{+} = g^{(2)}_{-} \neq g^{(2)}_{\pm}$.
Finally, if $q = n_{\mathrm{ch}} \neq 0$ we shall have that  $g^{(2)}_{+} \neq g^{(2)}_{-} \neq g^{(2)}_{\pm}$.
To infer the correlation properties of the multiparticle final state the multiplicity distribution of the positive and negative species
should be separately assessed and, eventually, cross-correlated to measure $g^{(2}_{\pm}$. 

The multiplicity distributions measured in hadronic reactions 
or even heavy ion collisions more often than not  
count all the charged species in the final state (not only the positively or negatively 
charged species). In analogy with the gravitational case, second-order (cross) correlations between positively and negatively charged distributions may 
represent a valid tool for the scrutiny of the multiparticle final state. Absent the latter 
measurements,  and only for purposes of comparison, it is therefore simpler to consider  
 a single mode of the field and to recall that
 \begin{equation}
0 \leq g^{(2)} -1  = \frac{D^2 - \langle \hat{N} \rangle}{\langle \hat{N} \rangle^2}\leq 1,
\label{MP}
\end{equation}
where $D^2=\langle\hat{N}^2\rangle - \langle\hat{N}\rangle^2$ is the 
dispersion. In Eq. (\ref{MP}) $g^{(2)}-1$  is defined as in Eq. (\ref{av4}). 
The expression in terms of the variance $D^2$ 
follows by recalling that $\langle \hat{a}^{\dagger}\,
\hat{a}^{\dagger} \,\hat{a}\,\hat{a}\rangle = \langle\hat{N}^2 \rangle - \langle \hat{N} \rangle$ where 
$\hat{N} = \hat{a}^{\dagger}\,\hat{a}$.  If $g^{(2)}=1$, $D^2= \langle\hat{N} \rangle $ as it happens 
for a standard coherent state characterized by a Poisson multiplicity distribution. 
According to the quantum optical terminology, if  
 $g^{(2)}>1$ the light is said to be bunched (with super-Poissonian field statistics)
  while if $g^{(2)} <1$ the light is said to be 
 antibunched (with sub-Poissonian field statistics) \cite{loudon2,mandel}. 
 In the case of a (single mode) Fock state $|n\rangle$ it can be readily 
 shown from the definition (\ref{MP})  that $g^{(2)}=1 - 1/n$. In more general terms the 
 degree of second order coherence of an arbitrary single-mode excitation must satisfy 
 $g^{(2)}\geq 1 -1/\langle \hat{N} \rangle$ where the equality is saturated in the case of a Fock state.
In short the gravitational multiplicity distributions derived in Eqs. (\ref{av7}), (\ref{av12})--(\ref{av13}) and (\ref{av15}) are characterized by pronounced tails exhibiting a degree of correlation which can be quantitatively assessed from the degree of second-order coherence $g^{(2)}$ ranging between $1$ and $2$.

The gravitational multiplicity distributions will now be compared and contrasted 
with the analog observables typical of hadronic collisions
(see \cite{feinberg,carruthers,gv2,grosse} for a collection of relevant review articles on multiple production in strong interactions).
Hadronic collisions at high energies lead to charged 
multiplicity distributions whose shapes are well fitted by a single negative binomial distribution in fixed intervals of central 
(pseudo)rapidity $\eta$ \cite{alice}. There are of course physical differences between 
the charged multiplicity distributions arising from purely leptonic 
initial states (such as electron-positron collisions) and purely hadronic 
initial states such as ($pp$ and $p\overline{p}$). The physical analysis of different reactions in a unified dynamical 
framework (such as, for instance, the clan model \cite{gv1})  would be interesting per se and it is anyway beyond the aims of the present investigation.  In what follows the attention will be limited, for practical reasons, to the recent $pp$
results at the LHC \cite{alice,atlas,cms}.
Denoting with the prime a derivation with respect to $z$ of the probability generating function of Eq. (\ref{av13b}), the mean and the dispersion 
of the distribution are given, respectively, by $\overline{n}= {\mathcal P}'(1)$ and 
by $D^2=  {\mathcal P}''(1) + {\mathcal P}'(1) - [{\mathcal P}'(1)]^2$ and, consequently,
\begin{equation}
g^{(2)} -1 = \frac{D^2}{\langle n\rangle^2} - \frac{1}{\langle n\rangle} = \frac{1}{k_{\mathrm{NB}}},\qquad C_{2} = g^{(2)} + \frac{1}{\overline{n}},
\label{h10}
\end{equation}
where $C_{q} = \langle n^{q}\rangle /\langle n\rangle^q$ are the normalized moments of the distribution and, in particular, 
$C_{2} = \langle n^2\rangle/\langle n\rangle^2$; the experimental collaborations instead of reporting the values 
of $k_{\mathrm{NB}}$ prefer to report sometimes the value of $C_{2}$ (see, e.g. second and third paper of \cite{alice}).
It is a property of the negative binomial distribution that all the higher-order moments (i.e. $C_{q}$ with $q\geq 2$) can all be expressed in terms of $\overline{n}$ and $k_{\mathrm{NB}}$. 
To swiftly outline the general features of the observed charged multiplicity distributions it is appropriate to look at the most 
recent measurements and, in particular, at the results obtained at the LHC \cite{alice,atlas,cms}. 
Consider in particular the data\footnote{We shall not dwell here on the details of the various experimental analyses such as the occurrence that the maximal multiplicity is always finite (i.e. $n$ never goes to infinity) and, in this sense, all the charged multiplicity distributions reconstructed from the experimental data should belong to a truncated set of multiplicities for fixed energy and in central (pseudo)rapidity intervals.} of the Alice collaboration \cite{alice} which are summarized in Tab. \ref{TABLE1} where 
the obtained results are also compared with the data of the UA5 experiment \cite{UA5} for $p\overline{p}$.
 \begin{table}[!ht]
\begin{center}
\begin{tabular}{|||c||c||c||c||c|||}
\hline
Data &  UA5\, $p\overline{p}$ \, $\sqrt{s} = 0.9\, \mathrm{TeV}$  &   Alice $pp$  $\sqrt{s} = 0.9\, \mathrm{TeV}$ & Alice $pp$  $\sqrt{s} = 2.36\, \mathrm{TeV}$ \\
\hline
\hline
$\langle N_{\mathrm{ch}} \rangle$ for $|\eta| < 0.5$ & 3.61 & 3.60& 4.47 \\
$k_{\mathrm{NB}}$ for $|\eta| < 0.5$ &  1.50& 1.46& 1.25\\
\hline
\hline
$\langle N_{\mathrm{ch}} \rangle$ for $|\eta| < 1.0$ & 7.38& 7.38& 9.08\\
$k_{\mathrm{NB}}$ for $|\eta| < 1.0$ & 1.62& 1.57& 1.37 \\
\hline
\hline
$\langle N_{\mathrm{ch}} \rangle$ for $|\eta| < 1.3$ & & 9.73& 11.86\\
$k_{\mathrm{NB}}$ for $|\eta| < 1.3$ &  no data & 1.67& 1.41\\
\hline
\hline
\end{tabular}
\caption{The parameters of the negative binomial distribution as they arise in different central (pseudo)rapidity intervals and different 
energies for the UA5 experiment and for the Alice experiment. The data are the ones published in \cite{alice} (see 
in particular the second paper).}
\label{TABLE1}
\end{center}
\end{table}
In Tab. \ref{TABLE1} $\langle N_{\mathrm{ch}} \rangle $ coincides with $\overline{n}$ of Eq. (\ref{h10}) (only central values are reported). By looking at Tab. \ref{TABLE1} three important points should be noticed.
For {\em fixed central (pseudo)rapidity intervals}, $k_{\mathrm{NB}}$ decreases as the centre of mass energy $\sqrt{s}$ increases while, as well known,  $\langle N_{\mathrm{ch}} \rangle$ increases  (logarithmically) with $\sqrt{s}$.
For {\em fixed centre of mass energy $\sqrt{s}$},  $k_{\mathrm{NB}}$ and $\langle N_{\mathrm{ch}} \rangle$ are both increasing with $|\eta|$, i.e. the absolute value of the (pseudo)rapidity $\eta$. The two preceding results are consistent with the trends already observed in the GeV energy domain (see, e.g. \cite{carruthers,gv2,grosse}). 

When $k_{\mathrm{NB}}$ decreases the tail of the distribution increases (i.e. correlations are present
for large multiplicities in comparison with the Poisson case).
There is a fourth point which is relevant for the forthcoming considerations: 
in the case of heavy ion collisions (see, e.g. \cite{phenix} for the cases of Au+Au and Cu + Cu up to energies $\sqrt{s} = 200 \,\mathrm{GeV}$) $k_{\mathrm{NB}}$ is  quite large (i.e. $1/k_{\mathrm{NB}}\to 0$, and, according to 
Eq. (\ref{h10}), $g^{(2)} \to 1$).  

By comparing the multiplicity distributions in strong interactions 
with their gravitational counterpart various interesting connections emerge and will now be swiftly summarized.
The inequality established in Eq. (\ref{MP}) for of gravitational multiplicity distributions 
holds also for hadronic multiplicity distributions in strong interactions (see Eq. (\ref{h10})); in both cases the Bose-Einstein distribution (i.e. 
$g^{(2)} \to 2$) and the Poisson distributions (i.e. $g^{(2)}\to1$) are, respectively, 
the signatures of {\em maximal and  minimal correlation} among the produced particles. 
The {\em maximally correlated} case in gravitational multiplicity distributions does not necessarily entail the presence 
of a mixed state but can even be the result of a pure state with zero charge. For {\em hadronic multiplicity distributions}, 
the decrease of $k_{\mathrm{NB}}$ (when the centre of mass energy of the collision increases and the central rapidity intervals are fixed)  is pointing towards the Bose-Einstein limit (i.e. $ k_{\mathrm{NB}}\to 1$). For the {\em gravitational multiplicity distributions} the {\em minimally correlated} case corresponds to the situation where the number of species present in the initial state dominates over the average number of produced pairs. 
For {\em hadronic multiplicity distributions} the {\em minimally correlated} case is expected in heavy ion collisions when 
$k_{\mathrm{NB}}$ is pretty large and the dispersion of the distribution roughly coincides 
with the average multiplicity (Poisson limit). While in the hadronic 
case $k_{\mathrm{NB}}$ can also be non-integer, in the gravitational case $k_{\mathrm{NB}}$ assumes only integer values and counts the charges of the initial state: this is the main difference between the hadronic and the gravitational multiplicity 
distributions. It is nonetheless remarkable that the distorted negative binomial distribution 
(deduced in Eq. (\ref{av12})) has a slight excess for low 
multiplicities while it is comparatively smaller than the standard negative binomial distribution for large multiplicities.
The gravitational multiplicity distributions, as their hadronic counterparts, all fall within the class 
of infinitely divisible distributions and they can be, in turn, classified in terms of the matrix elements of the positive 
(dscrete) series of $SU(1,1)$. In the case of gravitational multiplicity distributions there is an absolute asymptote 
for $1/k_{\mathrm{NB}}$, i.e. 
\begin{equation}
\frac{1}{k_{\mathrm{NB}}}\to 1, \qquad C_{2} \to 2 + \frac{1}{\langle N_{\mathrm{ch}}\rangle},
\end{equation}
where $C_{2}$ is the second normalized moment (see Eq. (\ref{h10}) and discussion thereafter). Assuming that the trend established at the LHC will be confirmed, it would be interesting to understand if such an asymptote is also present for hadronic multiplicity distributions (in central (pseudo)rapidity intervals) as the centre of mass energy in $pp$ collisions increases.

\end{document}